\documentclass[english,aps,prl,superscriptaddress,twocolumn,address,showpacs,showacknowledgments]{revtex4}
\pdfoutput=1
\usepackage[T1]{fontenc}
\usepackage[latin9]{inputenc}
\usepackage{babel}
\usepackage{amstext}
\usepackage{amssymb}
\usepackage{graphicx}
\usepackage[unicode=true,pdfusetitle,
 bookmarks=true,bookmarksnumbered=false,bookmarksopen=false,
 breaklinks=false,pdfborder={0 0 1},backref=false,colorlinks=false]
 {hyperref}

\makeatletter
\@ifundefined{textcolor}{}
{%
 \definecolor{BLACK}{gray}{0}
 \definecolor{WHITE}{gray}{1}
 \definecolor{RED}{rgb}{1,0,0}
 \definecolor{GREEN}{rgb}{0,1,0}
 \definecolor{BLUE}{rgb}{0,0,1}
 \definecolor{CYAN}{cmyk}{1,0,0,0}
 \definecolor{MAGENTA}{cmyk}{0,1,0,0}
 \definecolor{YELLOW}{cmyk}{0,0,1,0}
 }

\usepackage{bbold}
\renewcommand\[{\begin{equation}}
\renewcommand\]{\end{equation}} 

\renewenvironment{eqnarray*}{\eqnarray}{\endeqnarray}

\makeatother

\begin{document}

\title{Optomechanically Induced Transparency in the Nonlinear Quantum Regime}

\author{Andreas Kronwald}

\email{andreas.kronwald@physik.uni-erlangen.de}

\selectlanguage{english}%

\affiliation{Friedrich-Alexander-Universität Erlangen-Nürnberg, Staudtstr. 7,
D-91058 Erlangen, Germany}

\author{Florian Marquardt}

\affiliation{Friedrich-Alexander-Universität Erlangen-Nürnberg, Staudtstr. 7,
D-91058 Erlangen, Germany}

\affiliation{Max Planck Institute for the Science of Light, Günther-Scharowsky-Straße
1/Bau 24, D-91058 Erlangen, Germany}

\begin{abstract}
Optomechanical systems have been shown both theoretically and experimentally
to exhibit an analogon to atomic electromagnetically induced transparency,
with sharp transmission features that are controlled by a second laser
beam. Here we investigate these effects in the regime where the fundamental
nonlinear nature of the optomechanical interaction becomes important.
We demonstrate that pulsed transistor-like switching of transmission
still works even in this regime. We also show that optomechanically
induced transparency at the second mechanical sideband could be a
sensitive tool to see first indications of the nonlinear quantum nature
of the optomechanical interaction even for single-photon coupling
strengths significantly smaller than the cavity linewidth. 
\end{abstract}
\maketitle
\global\long\def\i{i}
\global\long\def\eps{\varepsilon}
\global\long\def\a{\hat{a}}
\global\long\def\b{\hat{b}}

\textit{Introduction \textendash{}} Optomechanics explores the coupling
between photons and phonons via radiation pressure. It aims at applications
in classical and quantum information processing as well as ultrasensitive
measurements and tests of fundamental quantum effects using mesoscopic
or macroscopic systems \cite{ANDP:ANDP201200226,Aspelmeyer:2013fk}.
Recently, a feature called {}``optomechanically induced transparency''
(OMIT) has been predicted theoretically \cite{PhysRevA.81.041803}
and observed experimentally \cite{Weis10122010,Teufel:2011fk,2011Safavi-Naeini:OMIT}:
The photon transmission through an optomechanical cavity is drastically
influenced when introducing a second laser beam. This leads to the
appearance of very sharp features in the transmission signal, which
can be controlled by the second beam. OMIT can thus be employed for
slowing and stopping light or for operating a {}``transistor'',
where photon transmission is switched on and off optically \cite{PhysRevA.81.041803,Weis10122010,2011Safavi-Naeini:OMIT,2011microwave_ampl}.
OMIT is an analogon of atomic electromagnetically induced transparency
\cite{RevModPhys.77.633}, where a medium consisting of three-level
atoms can be made transparent by illuminating it with a second laser. 

The optomechanical interaction is fundamentally nonlinear at the quantum
level. However, in most optomechanical systems the coupling between
single photons and phonons is small compared to dissipation rates
(except in cold atom clouds \cite{Murch:Granularity,Brennecke10102008,2012noncllight},
which have other constraints). It will therefore be extremely challenging
to detect effects of this nonlinearity on the quantum level. Indeed,
all the optomechanical quantum phenomena observed so far can be described
in a simpler linear model, where the coupling is effectively enhanced
via the photon number.

Nevertheless, experiments are recently making progress in increasing
the coupling strength \cite{Teufel_Strong_Coupling,Chan_strong_coupling,Verhagen_strong_coupling,chan:081115},
coming closer to the nonlinear quantum regime. That regime has attracted
large theoretical interest leading to the prediction of optical Schrödinger
cat states \cite{PhysRevA.55.3042,PhysRevA.56.4175}, a classical
to quantum crossover in nonlinear optomechanical systems \cite{2008_NJP_OMinstab_MaxLudwig},
non-Gaussian \cite{PhysRevLett.107.063602} and non-classical mechanical
states \cite{PhysRevLett.109.253601,Xu:2012uq,1367-2630-15-3-033023},
as well as multiple cooling resonances \cite{PhysRevA.85.051803}.
Certain dark states \cite{PhysRevA.87.053849}, photon antibunching
\cite{PhysRevLett.107.063601,PhysRevA.87.013847}, a crossover from
sub-Poissonian to super-Poissonian statistics and photon cascades
\cite{PhysRevA.87.013847} may be observed. Two-mode setups \cite{PhysRevLett.109.013603,PhysRevLett.109.063601}
and collective effects in optomechanical arrays \cite{PhysRevLett.109.223601}
have been shown to be favorable for reaching strong quantum nonlinearities.
\begin{figure}[t]
\includegraphics{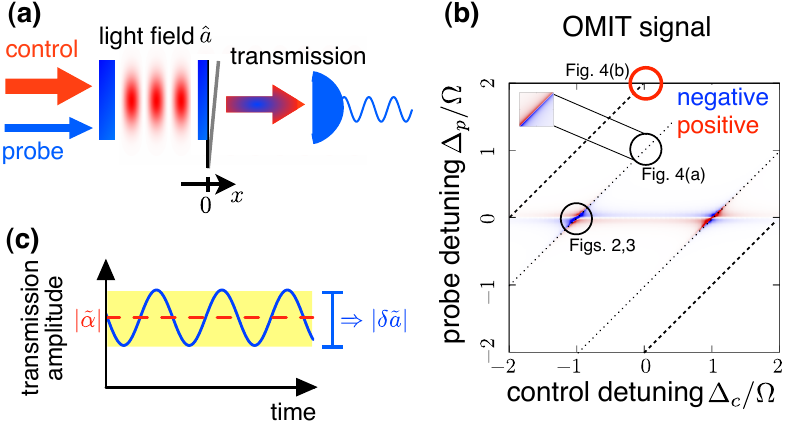}

\caption{(a) Standard optomechanical setup driven by a control and a probe
laser. (b) Classical expectation of the OMIT signal as a function
of the probe and control detuning, cf. Eq. (\ref{eq:OMIT_linearized}).
If the beat frequency of the two lasers $|\omega_{c}-\omega_{p}|\sim\Omega$,
a signal is expected (dotted, black lines). We study the OMIT signal
at the circles. In particular, we find an OMIT signal at the second
mechanical sideband (red, thick circle) even at moderate single-photon
coupling, where a classical analysis fails to show the signal. This
is a clear signature of the optomechanical quantum nonlinearity. (c)
Field amplitude in a frame rotating at the control laser frequency
$\omega_{c}$. The control generates a constant transmission amplitude
$\tilde{\alpha}$ (red, dashed line). The probe induces oscillations
at the beat frequency $\omega_{p}-\omega_{c}$. The amplitude of these
oscillations, Eq. (\ref{eq:a_exp}), is our signal. {[}Parameters:
$g=g_{0}\left|\alpha\right|=0.08\,\Omega$ , $\kappa=\Omega/8$, $\Gamma_{M}=0.01\,\Omega${]}.}

\label{fig:FIG_1}
\end{figure}

In this paper, we go beyond the classical analysis of OMIT \cite{PhysRevA.81.041803,ediss10940,Weis10122010,PhysRevA.86.013815}
and analyze OMIT in the nonlinear quantum regime. By simulations of
the full quantum dissipative dynamics, we study the spectroscopic
signal and the time-evolution during pulsed operation. In analyzing
OMIT at mechanical sidebands, we find that OMIT could be a crucial
tool to observe first tell-tale effects of the nonlinear quantum regime
even for single-photon coupling strengths much smaller than the cavity
linewidth, in contrast to the standard single-beam situation \cite{PhysRevLett.107.063601,PhysRevLett.107.063602}. 

\textit{Model \textendash{}} We consider a generic optomechanical
system, where an optical cavity is coupled to mechanical motion, cf.
Fig. \ref{fig:FIG_1}(a). The system's Hamiltonian reads \cite{PhysRevA.51.2537}

\begin{equation}
\hat{H}=\hbar\omega_{\text{cav}}\hat{a}^{\dagger}\hat{a}+\hbar\Omega\hat{b}^{\dagger}\hat{b}-\hbar g_{0}\left(\hat{b}^{\dagger}+\hat{b}\right)\hat{a}^{\dagger}\hat{a}+\hat{H}_{\text{dr}}\,,\label{eq:standard_OM_Hamiltonian}
\end{equation}
where $\hat{a}$ ($\hat{b}$) is the photon (phonon) annihilation
operator, $\omega_{\text{cav}}$ ($\Omega$) the cavity (mechanical)
resonance frequency, and $g_{0}$ the optomechanical coupling between
single photons and phonons. $\hat{H}_{\text{dr}}$ describes the two-tone
driving, 
\begin{equation}
\hat{H}_{\text{dr}}=\hbar\left[\varepsilon_{c}e^{-\mathrm{i}\omega_{c}t}+\eps_{p}e^{-\mathrm{i}\omega_{p}t}\right]\hat{a}^{\dagger}+\text{H.c}\,,
\end{equation}
where $\omega_{c}$ ($\omega_{p}$) and $\eps_{c}$ ($\eps_{p}$)
are the control (probe) laser frequency and amplitude, respectively.
In order to neglect other mechanical resonances, one has to assume
$g_{0}\ll\Omega$. We also assume near-resonant excitation (and narrow-band
detection) of the mechanical sidebands under consideration.

The optomechanical interaction can be diagonalized (for $\eps_{c,p}=0$)
by shifting the mechanical equilibrium position by $\delta x\propto2n_{a}g_{0}/\Omega$
depending on the photon number $n_{a}$ ({}``polaron transformation'')
\cite{PhysRevA.55.3042,PhysRevA.56.4175}. This will allow us to understand
OMIT in terms of interference pathways in the resulting level scheme.
The corresponding eigenstates read $\left|n_{a},n_{b}\right\rangle $
where $n_{a}$ ($n_{b}$) is the number of photons (phonons in the
shifted frame). The eigenenergies read $E\left(n_{a},n_{b}\right)/\hbar=\omega_{\text{cav}}^{\text{eff}}n_{a}+\Omega n_{b}-g_{0}^{2}n_{a}\left(n_{a}-1\right)/\Omega$
\cite{PhysRevA.55.3042,PhysRevA.56.4175}, where $\omega_{\text{cav}}^{\text{eff}}=\omega_{\text{cav}}-g_{0}^{2}/\Omega$
is the effective cavity resonance frequency. The control/probe detuning
from $\omega_{\text{cav}}^{\text{eff }}$ is defined as $\Delta_{c/p}=\omega_{c/p}-\omega_{\text{cav}}^{\text{eff}}$. 

\textit{Dissipative Dynamics \textendash{}} The dissipative dynamics
for weak optomechanical coupling can be described by a Lindblad master
equation

\begin{equation}
\dot{\hat{\rho}}=\frac{\mathrm{i}}{\hbar}\left[\hat{\rho},\hat{H}\right]+\kappa\mathcal{D}[\hat{a}]\hat{\rho}+\Gamma_{M}\left(n_{\text{th}}+1\right)\mathcal{D}[\hat{b}]\hat{\rho}+\Gamma_{M}n_{\text{th}}\mathcal{D}[\hat{b}^{\dagger}]\hat{\rho}\,.\label{eq:Lindblad}
\end{equation}
$\hat{\rho}$ denotes the density matrix for the optical and mechanical
mode. $\kappa$ is the photon loss rate (due to loss through both,
input \textit{and} output mirror), and $\Gamma_{M}$ the phonon decay
rate. $n_{\text{th}}$ is the thermal occupancy of the mechanical
bath and $\mathcal{D}\left[\hat{A}\right]\hat{\rho}=\hat{A}\hat{\rho}\hat{A}^{\dagger}-\hat{A}^{\dagger}\hat{A}\hat{\rho}/2-\hat{\rho}\hat{A}^{\dagger}\hat{A}/2$
is the Lindblad dissipation superoperator. We solve (\ref{eq:Lindblad})
in the time-domain numerically. This allows to also consider pulse-based
schemes. 

\textit{Two-tone transmission \textendash{} }In the steady state,
the intracavity field amplitude in a frame rotating at the control
frequency $\omega_{c}$ is defined by
\begin{equation}
\left\langle \a\right\rangle =\alpha+\eps_{p}\left[\delta a_{1}e^{-\i\left(\omega_{p}-\omega_{c}\right)t}+\delta a_{-1}e^{\i\left(\omega_{p}-\omega_{c}\right)t}\right]\,.\label{eq:a_exp}
\end{equation}
The control beam induces a constant amplitude $\alpha$, whereas $\delta a_{\pm1}$
are two (first-order) sidebands due to the probe. Higher harmonics
of the beat frequency $\omega_{p}-\omega_{c}$ are also present due
to the nonlinear interaction. However, they are weak in our analysis
and have been omitted in (\ref{eq:a_exp}). 

The experimentally accessible transmitted field amplitude $\left\langle \a_{\text{out}}\right\rangle $
is related to the cavity field (\ref{eq:a_exp}) by the input-output
relation $\left\langle \a_{\text{out}}\right\rangle =\sqrt{\kappa_{O}}\left\langle \a\right\rangle $
\cite{2008_ClerkDevoretGirvinFMSchoelkopf_RMP}, where $\kappa_{O}$
is the output mirror decay rate. In the following, we analyze what
we term the normalized probe beam transmission 
\[
\left|\delta\tilde{a}\right|^{2}=\kappa_{O}\eps_{p}^{2}\left(\left|\delta a_{1}\right|^{2}+\left|\delta a_{-1}\right|^{2}\right)/\left|\delta a_{\text{out}}^{\text{max}}\right|^{2}\,.
\]
This is essentially the intensity transmitted at the probe beam frequency
$\omega_{p}$ and at the other sideband, $2\omega_{c}-\omega_{p}$,
divided by the incoming probe intensity $\left|\delta a_{\text{out}}^{\text{max}}\right|^{2}=\kappa_{O}\cdot4\eps_{p}^{2}/\kappa^{2}$.
$\left|\delta a_{\pm1}\right|$ can be measured via heterodyning \cite{Gardiner_Zoller},
i.e. mixing $\a_{\text{out}}$ with a local oscillator at $\omega_{c}$
and obtaining the power in the signal at $\omega_{p}\pm\omega_{c}$. 

To isolate signatures in the probe beam transmission which emerge
due to the presence of the control beam, we introduce the {}``OMIT
signal''. It is defined as the difference of $\left|\delta\tilde{a}\right|^{2}$
with and without control laser.

\textit{Standard prediction \textendash{}} OMIT has so far been studied
only in the \textit{linearized} regime of optomechanics, where the
quantum nonlinearity is neglected. In this regime, the OMIT signal
depends only on the product $g_{0}|\alpha|$ of the coupling and the
intra-cavity field amplitude $\alpha=\i\eps_{c}/\left(\i\Delta_{c}-\kappa/2\right)$.
For this to be valid, $g_{0}\ll\kappa$.

\begin{figure}[t]
\includegraphics{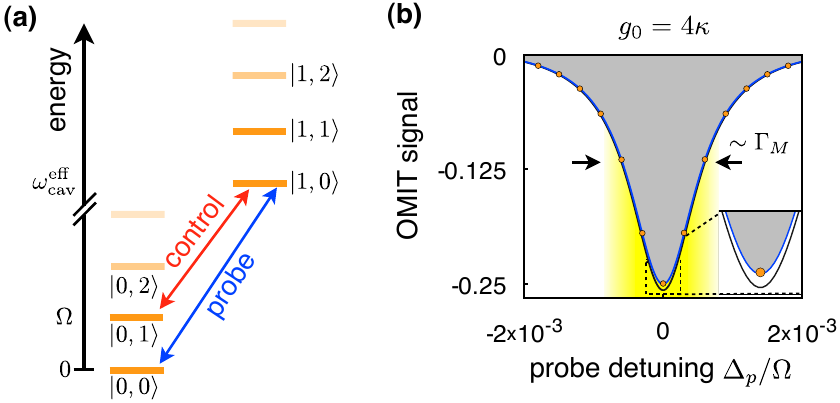}

\caption{Quantum nonlinearities and OMIT. (a) Energy level scheme of the optomechanical
system with levels $\left|n_{a},n_{b}\right\rangle $, cf. main text.
$n_{a}$ ($n_{b}$) denotes the number of photons (phonons). Here,
for example, the probe couples $\left|0,0\right\rangle \leftrightarrow\left|1,0\right\rangle $
since the detuning $\Delta_{p}=\omega_{p}-\omega_{\text{cav}}^{\text{eff}}=0$.
The control couples $\left|0,1\right\rangle \leftrightarrow\left|1,0\right\rangle $
since $\Delta_{c}=\omega_{c}-\omega_{\text{cav}}^{\text{eff}}=-\Omega$.
(b) The OMIT signal. Orange circles: Numerical results for $g_{0}/\kappa=4$.
Black line: Expectation for the standard, classical regime, Eq. (\ref{eq:OMIT_linearized}).
Blue line: Expectation of Eq. (\ref{eq:OMIT_linearized}), but including
Franck-Condon factors and $\left|\alpha\right|^{2}\to\left\langle \hat{a}^{\dagger}\hat{a}\right\rangle $,
see main text. {[}Parameters: $\kappa=\Omega/40$, $\Gamma_{M}=10^{-3}\,\Omega$,
$\eps_{c}=10^{-2}\,\Omega$, $\Delta_{c}=-\Omega$, $n_{\text{th}}=0${]}.}

\label{fig:FIG_2}
\end{figure}

We recall that a common OMIT signature arises when the control laser
drives the cavity on the red sideband, i.e. $\Delta_{c}=-\Omega$.
The probe beam transmission as a function of the probe detuning $\Delta_{p}$
then shows a transmission dip on resonance (i.e. $\Delta_{p}=0$),
cf. Fig. \ref{fig:FIG_2}(b). The dip's width is $\Gamma_{M}+\Gamma_{\text{opt}}\ll\kappa$
(with $\Gamma_{\text{opt}}=4g_{0}^{2}\left|\alpha\right|^{2}/\kappa$).
The normalized probe beam transmission reads \cite{PhysRevA.81.041803,ediss10940}
\begin{equation}
\left|\delta\tilde{a}\right|^{2}=\frac{\kappa^{2}}{4}\left|\frac{1}{-\i\Delta_{p}+\frac{\kappa}{2}-2\i\frac{g_{0}^{2}}{\Omega}\left|\alpha\right|^{2}\chi\left[\omega_{p}-\omega_{c}\right]}\right|^{2}\,,\label{eq:OMIT_linearized}
\end{equation}
where $\chi^{-1}\left[\omega\right]=1-\left(\omega/\Omega\right)^{2}-\i\omega\Gamma_{M}/\Omega^{2}$
is the (rescaled) mechanical susceptibility.

If $\Delta_{p}\approx0$, the beat frequency $\omega_{p}-\omega_{c}$
between probe and control is given by the mechanical frequency $\Omega$.
Thus, the mechanical resonator is driven by a force oscillating at
its eigenfrequency and the resonator starts to oscillate coherently.
This motion induces sidebands on the cavity field, generating photons
with frequency $\omega_{p}$. These interfere destructively with the
probe beam, leading to a transmission dip. Typically, OMIT has been
studied in a regime where $\Gamma_{M}\ll\Gamma_{\text{opt}}$ and
$g_{0}\left|\alpha\right|\ll\kappa\ll\Omega$ \cite{Weis10122010,Teufel:2011fk,2011Safavi-Naeini:OMIT},
such that the OMIT dip's width is $\sim\Gamma_{\text{opt}}$. 

We now focus on the regime where the quantum nonlinearity becomes
important. Quantum nonlinear features can be unambiguously distinguished
from classical effects (or linear quantum effects) by studying their
dependence on the {}``quantum parameter'' $g_{0}/\kappa$ \cite{Murch:Granularity,2008_NJP_OMinstab_MaxLudwig}.
Let us imagine that Planck's constant $\hbar\to0$. In this limit,
all classical effects remain while all quantum effects become vanishingly
small. In the context of optomechanics, varying $\hbar$ is equivalent
to keeping all classical parameters ($\Delta,\mbox{\ensuremath{\kappa},\ensuremath{\ldots})}$
fixed while tuning the {}``quantum parameter'' $g_{0}/\kappa\propto\sqrt{\hbar}$
\cite{2008_NJP_OMinstab_MaxLudwig,Murch:Granularity,Aspelmeyer:2013fk}.
As $g_{0}\to0$, we increase the laser power $\eps_{c}$, hence $g_{0}\left|\alpha\right|=\text{const}$.
This retains the size of the classical OMIT signal, cf. (\ref{eq:OMIT_linearized}).
In contrast, any truly quantum-mechanical nonlinear effects vanish
as $g_{0}\to0$. Note that for our parameters, a \textit{single} photon
can have a large impact \cite{PhysRevLett.107.063601,PhysRevLett.107.063602},
so we limit ourselves to weak laser driving, i.e. $\eps_{p},\eps_{c}\ll\kappa$
(thus $\left|\alpha\right|\ll1$) in contrast to the standard OMIT
scenario, where $\epsilon_{p,c}$ can be arbitrary. This also implies
that the OMIT dip's width is $\sim\Gamma_{M}$.

\begin{figure}
\includegraphics{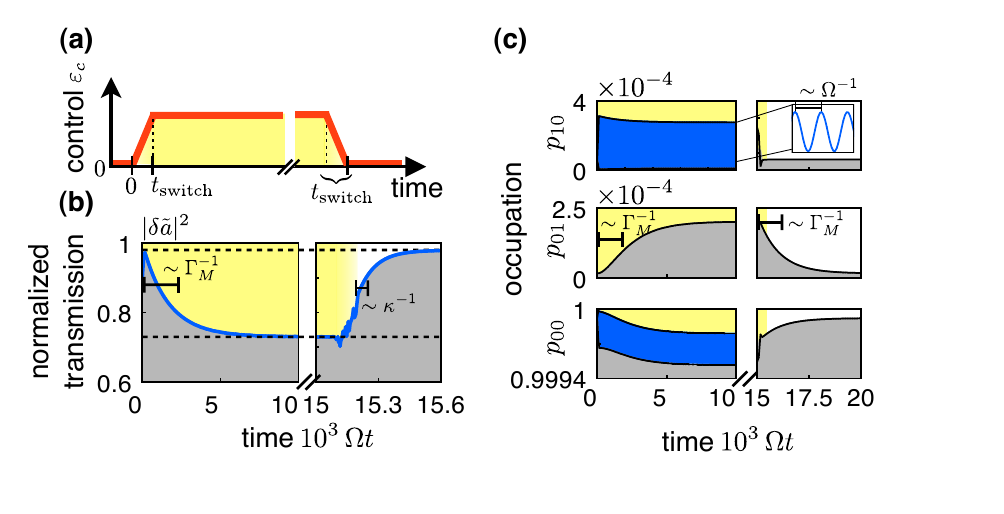}\caption{Optomechanical transistor for large $g_{0}/\kappa$. (a) The probe
drives the cavity continuously. At time $t=0$, the control laser
power is ramped-up linearly until $\Omega t_{\text{switch}}\gg1$.
This decreases the normalized probe beam transmission on a scale $\sim\Gamma_{M}^{-1}$,
cf. (b). When switching off the control linearly, $\left|\delta\tilde{a}\right|^{2}$
increases on a scale $\sim\kappa^{-1}\ll\Gamma_{M}^{-1}$. (c) Occupation
transfer between individual quantum states induced by the control.
Oscillations at $\Omega$ are clearly visible in $p_{10}$ (i.e. for
the state with 1 photon, 0 phonons). {[}Parameters: same as in Fig.
\ref{fig:FIG_2}, $\Delta_{p}=0$, $\Omega t_{\text{switch}}=100${]}.}

\label{FIG_2_1}
\end{figure}

\textit{Main OMIT dip \textendash{}} To compare against \textit{classical}
OMIT predictions, we first focus on the OMIT dip at resonance, while
keeping the full nonlinear \textit{quantum} interaction of (\ref{eq:standard_OM_Hamiltonian}),
cf. Fig. \ref{fig:FIG_2}(b). Consider the level scheme of Fig.~\ref{fig:FIG_2}(a).
Since both lasers are assumed to be weak, only the zero and one photon
ladders are important. We again assume $\Delta_{c}=-\Omega$, such
that the control hybridizes the states $|0,1\rangle\leftrightarrow|1,0\rangle$.
This leads to a destructive interference of the two probe excitation
pathways at $\Delta_{p}=0$ and thus the OMIT dip. 

The most important change in the OMIT signal is that the main OMIT
dip's depth is modified due to shifted phonon states: The probe laser
drives the transition $\left|0,0\right\rangle \leftrightarrow\left|1,0\right\rangle $
resonantly, i.e. the transition of a $0$-phonon state and a \textit{shifted}
$1$-phonon state. This leads to the Franck-Condon factor $\exp[-\left(g_{0}/\Omega\right)^{2}]$
which will enter the numerator of (\ref{eq:OMIT_linearized}) \cite{PhysRevLett.107.063602}.
Also, the photon number $\left\langle \hat{a}^{\dagger}\hat{a}\right\rangle $
($\left|\alpha\right|^{2}$ in the classical theory) entering the
denominator of (\ref{eq:OMIT_linearized}) is changed in this regime
\cite{PhysRevLett.107.063602}. When taking these modifications into
account in (\ref{eq:OMIT_linearized}) we obtain quantitative agreement,
cf. Fig. \ref{fig:FIG_2}(b).

\begin{figure}[t]
\includegraphics{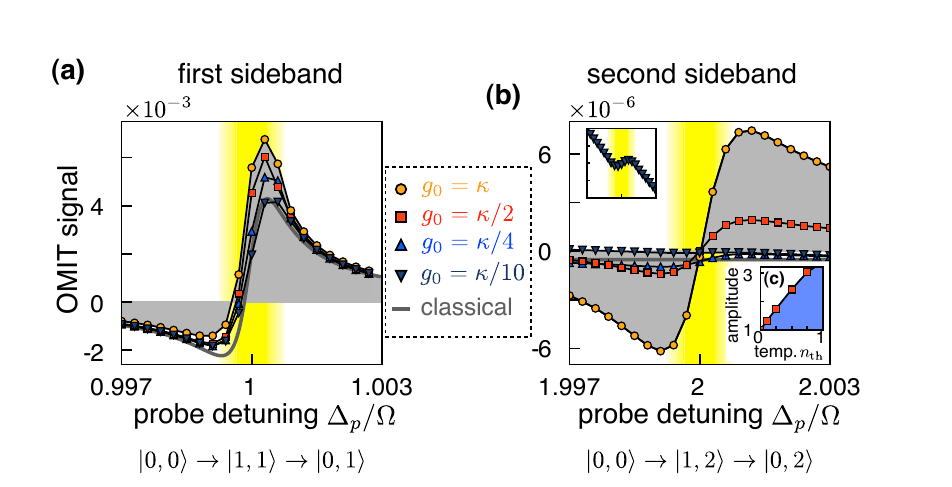}

\caption{Quantum to classical crossover. The OMIT signal at the second sideband
(b) vanishes in the classical limit $g_{0}/\kappa\to0$, hence being
a clear signature of the quantum nonlinearity. It is visible even
for moderate coupling strengths $g_{0}<\kappa$. The yellow regions
indicate the OMIT features's width $\sim\Gamma_{M}$. (a) Since $\Delta_{p}=\Omega$
and $\Delta_{c}=0$, probe and control couple the levels $\left|0,0\right\rangle \leftrightarrow\left|1,1\right\rangle $
and $\left|0,1\right\rangle \leftrightarrow\left|1,1\right\rangle $
respectively. (b) Since $\Delta_{p}=2\Omega$ and $\Delta_{c}=0$,
$\left|0,0\right\rangle \leftrightarrow\left|1,2\right\rangle $ and
$\left|0,2\right\rangle \leftrightarrow\left|1,2\right\rangle $ are
coupled, respectively. Symbols: OMIT signal for different $g_{0}/\kappa$
where $g_{0}\left|\alpha\right|$ and $\kappa$ are kept fixed. Grey
line: classical expectation, cf. (\ref{eq:OMIT_linearized}). Inset
of (b): OMIT signal for $g_{0}/\kappa=1/10$. The axis are the same
as in (b). The OMIT signal varies in a range $\sim10^{-7}$. Note
that the signal strength can be increased by increasing $g_{0}\left|\alpha\right|$.
(c) Amplitude of the Fano resonance at the second sideband (i.e. the
difference between the maximum and minimum) versus the thermal phonon
number $n_{\text{th}}$ (normalized to the amplitude at $n_{t\text{h}}=0$).
{[}Parameters: $\kappa=\Omega/8$, $\Gamma_{M}=10^{-3}\,\Omega$,
$\Delta_{c}=0$, $n_{\text{th}}=0$, $g_{0}\eps_{c}=1.25\cdot10^{-3}\,\Omega^{2}=\text{const}$.
(This value has been chosen to keep the Hilbert space manageable as
$g_{0}/\kappa\rightarrow0$). (c): Same as in (b), $g_{0}/\kappa=1/2${]} }

\label{FIG:3}
\end{figure}

Thus, the standard OMIT dip in the nonlinear quantum regime is still
controlled by the photon number $\left\langle \a^{\dagger}\a\right\rangle $
only, allowing the operation of an optomechanical transistor in a
pulsed scheme. 

\textit{Optomechanical transistor \textendash{}} Let us consider a
resonant probe beam. At $t=0$, we ramp-up the control power linearly
(to avoid spurious transients). At $\Omega t_{\text{switch}}\gg1$,
the red detuned control ($\Delta_{c}=-\Omega)$ reaches its maximum
power, cf. Fig. \ref{FIG_2_1}(a). This decreases the probe beam transmission
$\left|\delta\tilde{a}\right|^{2}$ on a timescale set by the OMIT
dip's width $\sim\Gamma_{M}^{-1}$. When reaching the steady state,
the control beam is linearly switched off. Then, $\left|\delta\tilde{a}\right|^{2}$
increases rapidly on a scale $\sim\kappa^{-1}$, cf. Fig. \ref{FIG_2_1}(b),
because control photons decay out of the cavity on this timescale.
The influence of the control can also be seen in the population $p_{n_{a},n_{b}}$
of the states $|n_{a},n_{b}\rangle$, cf. Fig. \ref{FIG_2_1}(c).
Control and probe together first increase the population of the one-photon
state $|1,0\rangle$. Then, on a scale $\sim\Gamma_{M}^{-1}$, the
one-phonon state's population $p_{0,1}$ increases, thus $\left|\delta\tilde{a}\right|^{2}$
decreases. Furthermore, population between the zero and one photon
ladder is exchanged coherently, leading to oscillations at $\omega_{p}-\omega_{c}=\Omega$.

\textit{Quantum to classical crossover \textendash{}} We now discuss
how signatures of the quantum nonlinearity could be observed using
OMIT even if the coupling strength $g_{0}<\kappa$, which is relevant
for present experiments. As discussed above, to distinguish nonlinear
quantum effects from classical effects, we study the OMIT signal as
a function of the quantum parameter $g_{0}/\kappa$ while keeping
the classical prediction (\ref{eq:OMIT_linearized}) unchanged.

The most significant signature of the quantum nonlinearity can be
obtained for a resonant control beam, i.e. $\Delta_{c}=0$ and observing
the OMIT signal close to the mechanical sidebands, i.e. $\Delta_{p}\approx n\Omega$,
where $n=1,2,\ldots$.

At the first mechanical sideband ($n=1$), the classical theory (\ref{eq:OMIT_linearized})
predicts a Fano-resonance, cf. Fig. \ref{FIG:3}(a). Fano resonances
in general have recently been discussed in the context of optomechanics
\cite{PhysRevLett.102.067202,PhysRevLett.102.207209,PhysRevLett.107.213604,1367-2630-15-4-045017,PhysRevA.87.063813}.
The Fano resonance emerges because the probe beam probes both the
first order, off-resonant $|1,0\rangle\leftrightarrow|0,0\rangle$
transition plus the second-order, resonant transition $|0,1\rangle\leftrightarrow|0,0\rangle$,
the latter being a joint effect of the probe and control. For small
quantum parameters $g_{0}\ll\kappa$, the OMIT signal converges to
the classical expectation. Upon increasing $g_{0}/\kappa$, the Fano
resonance becomes slightly more pronounced as the relevant $\left|0,0\right\rangle \leftrightarrow\left|1,1\right\rangle $
transition becomes more likely due to the Franck Condon factor $\sim\left(g_{0}/\Omega\right)$
(in leading order). Note that the strength of the OMIT signal can
be increased by increasing $g_{0}|\alpha|$.

\textit{Second sideband OMIT as a sensitive probe \textendash{}} Let
us consider the second mechanical sideband, $\Delta_{p}\approx2\Omega$,
while, again, $\Delta_{c}=0$, cf. Fig. \ref{FIG:3}(b). The classical
analysis (whether linearized or fully nonlinear \cite{PhysRevA.86.013815})
fails to show an OMIT signature here, because it does not capture
transitions to sidebands $n>1$. However, when including the quantum
nonlinearity, OMIT signatures do exist. These vanish in the classical
limit $g_{0}/\kappa\to0$ and are thus solely due to the quantum nonlinearity. 

This feature emerges due to the two-photon transition $|0,0\rangle\mapsto|1,2\rangle\mapsto|0,2\rangle$,
with additional interference of off-resonant transitions probed by
the probe beam. The two-photon transition is enabled due to shifted
phonon states. Importantly, we observe this significant feature even
for moderate coupling $g_{0}=\kappa/10$ (in contrast to other quantum
signatures which require $g_{0}>\kappa$ \cite{PhysRevLett.107.063601,PhysRevLett.107.063602}).
The OMIT signal increases with increasing $g_{0}/\kappa$. This is
because the relevant higher order sideband transition rates increase
due to an increase of the Franck-Condon factors with $g_{0}$. Thus,
we predict that a two-tone driving experiment should be able to identify
quantum signatures of the optomechanical interaction even for moderate
single-photon coupling strengths $g_{0}<\kappa\ll\Omega$. 

We now discuss the influence of temperature on the OMIT signal at
the second sideband, cf. Fig. \ref{FIG:3}(c). As we increase temperature,
we find that the Fano resonance amplitude even increases, while the
dependence on $g_{0}/\kappa$ indicates that the effect is still a
signature of quantum nonlinearities, vanishing in the classical limit.
A possible explanation is that higher phonon states are now thermally
occupied. Thus, the transitions $\left|0,n_{b}\right\rangle \leftrightarrow\left|1,n_{b}+2\right\rangle $
with $n_{b}>0$ are additionally probed. The corresponding Franck-Condon
factors increase with $n_{b}$ (if $g_{0}\ll\Omega$), leading to
the observed enhancement. 

We now discuss what happens as we increase the probe driving strength
$\eps_{p}$. We find that even for $\eps_{p}\sim\eps_{c}$ the OMIT
signal does not change. The probe beam drives the cavity on the second
sideband and thus the number of probe photons inside the cavity is
still much lower than the number of control photons (as long as $\eps_{p}/\eps_{c}\ll\Omega/\kappa$)
- the probe is still a small perturbation. This is experimentally
relevant, since one can therefore increase the absolute probe transmission
by increasing the probe intensity. This also holds for the first mechanical
sideband (but not for the main OMIT dip, where the OMIT signal begins
to be suppressed even for $\eps_{p}/\eps_{c}\sim\kappa/\Omega$). 

\textit{\emph{We conclude that the challenge to bring out nonlinear
quantum signatures in optomechanical systems will be greatly aided
by two-tone driving.}}

\textit{Acknowledgments \textendash{}} We acknowledge fruitful discussions
with Max Ludwig, Vittorio Peano, Steven Habraken, and Aashish Clerk.
Financial support from the DARPA ORCHID, the Emmy-Noether program,
the European Research Council and the ITN cQOM is gratefully acknowledged. 

Note added: While preparing this manuscript for upload, two related
works appeared on the arXiv \cite{PhysRevLett.111.053602,PhysRevLett.111.053603}.

\bibliographystyle{apsrev4-1}
\bibliography{Literature}

\end{document}